\documentclass[12pt]{article}
\usepackage{times}
\usepackage{hyperref}

\usepackage{latexsym}

\DeclareMathAlphabet{\doba}{U}{msb}{m}{n}
\gdef\mC{\doba{C}}

\gdef\mR{\doba{R}}

\def\qed{{\leavevmode\unskip\nobreak\hfil\penalty 50\hskip 1em%
  \hbox{}\nobreak\hfil\lower 1pt\hbox{$\Box$\kern-.5pt}\parfillskip 0pt
  \finalhyphendemerits 0\par\bigbreak}}
\def\qedmath#1{\setbox0\hbox{$\displaystyle #1$}\templaenge=\textwidth\advance\templaenge by -\wd0%
\setbox1\hbox{$\Box$}\advance\templaenge by -2\wd1%
$$#1\hbox to0pt{\kern.5\templaenge$\Box$\kern-.5pt\hss}$$\par\bigbreak}

\def\la{{\lambda}}

\def\ga{{\gamma}}
\def\ep{{\varepsilon}}

\def\ph{{\varphi}}

\def\ze{{\zeta}}

\def\del{{\partial}}

\def\cD{\mathcal{D}}
\def\cE{\mathcal{E}}

\def\cH{\mathcal{H}}

\def\cO{\mathcal{O}}

\DeclareMathAlphabet{\goth}{U}{euf}{m}{n}

\def\dann{\Rightarrow}

\def\<{\langle}
\def\>{\rangle}

\def\x{\times}

\def\ie{i.\thinspace e.\ \ignorespaces}
\def\eg{e.\thinspace g.\ \ignorespaces}

\def\supp{\mathop{{\rm supp}}}
\def\dist{\mathop{{\rm dist}}}

\def\Id{\mathop{{\rm Id}}}

\def\ddt{\frac{d}{dt}}
\def\delt{\frac{\partial}{\partial t}}
\def\inv{^{-1}}

\def\mfd{\mbox{manifold }}
\def\cpt{compact }

\def\cnct{connected }

\def\Rm{Riemannian }

\newcommand{\nc}{\newcommand}

\nc{\pf}{\mbox{\em Proof. }}
\nc{\dfn}{\mbox{\bf Definition }}
\nc{\ex}{\mbox{\bf Example }}
\nc{\rmk}{\mbox{\bf Remark }}
\nc{\pbl}{\mbox{\bf Problem }}
\nc{\qes}{\mbox{\bf Question }}

\begin{document}

\title{Semi-Bounded Restrictions of Dirac Type Operators and the Unique
       Continuation Property}
\author{Christian B\"ar and Alexander Strohmaier}
\date{9.8.2000}
\maketitle

\begin{abstract}
\noindent Let $M$ be a connected Riemannian \mfd and let $D$ be a
Dirac type operator acting on smooth compactly supported sections
in a Hermitian vector bundle over $M$. Suppose $D$ has a
self-adjoint extension $A$ in the Hilbert space of
square-integrable sections. We show that any $L^2$-section $\ph$
contained in a closed $A$-invariant subspace onto which the
restriction of $A$ is semi-bounded has the unique continuation
property: if $\ph$ vanishes on a non-empty open subset of $M$,
then it vanishes on all of $M$.
\end{abstract}





\section{Introduction}

There is a large class of unique continuation theorems for
solutions to second order differential equations. For solutions to
Dirac type equations on Riemannian manifolds there exists a number
of methods to show the weak unique continuation property for
solutions, \ie a solution vanishing on a non-void open set
vanishes identically (\cite{booss99ppa}).

The relativistic quantum theory of a single electron is described
by the Hilbert space of 4-component spinors
$\mathcal{H}:=L^2(\mR^3,\mC^4)$ and the Dirac operator $H_0$ for
mass $m>0$, which is essentially self-adjoint on the dense
subspace $C^\infty_c(\mR^3,\mC^4)$. If $P_+$ and $P_-$ are the
orthogonal projections onto the spectral subspaces for the
positive and negative part of the spectrum of the Dirac operator,
the physical state space of the theory is the space of rays in the
Hilbert space $\mathcal{H}_+:=P_+ \mathcal{H}$. The time evolution
is given by the unitary group generated by the Dirac operator
restricted to $\mathcal{H}_+$. For such a physical state $\psi \in
\mathcal{H}_+, \Vert \psi \Vert=1$, and any region $\mathcal{O}
\subset \mR^3$ the integral $\int_\mathcal{O} \<\psi(x),\psi(x) \>
d^3x$ is the probability to find the electron in $\mathcal{O}$. It
is known for the case of the free Dirac operator that functions in
$\mathcal{H}_+$ have the weak unique continuation property, see
\cite[Corollary 1.7]{thaller92a}. Physically this means, that one
cannot localize a single electron in a bounded region. In the
fully quantized theory of the electron field this property has
important consequences. The reason for this is that the
quantization procedure requires the splitting of $\mathcal{H}$
into a direct sum $\mathcal{H}_+ \oplus \mathcal{H}_-$, see
\cite[Section 10.1]{thaller92a} for details. The weak continuation
property implies however that $P_+ C^\infty_c(\mathcal{O},\mC^4)$
is already dense in $\mathcal{H}_+$ for any non-void open set
$\mathcal{O} \subset \mR^3$. Hence the local information is
destroyed by the splitting. As a consequence the vacuum vector
turns out to be cyclic for the local field algebras. This
Reeh-Schlieder property of the vacuum is a general property of
quantum field theories in Minkowski space-time and can be proved
in the Wightman framework even for interacting fields
(\cite{reeh-schlieder61a}).

Replacing $\mR^3$ by an arbitrary Riemannian manifold $(M,g)$ it
is natural to ask whether the weak unique continuation property
still holds not only for solutions to Dirac type equations but
also, as in the case of $\mR^3$, for elements of certain spectral
subspaces. In this note we will prove that such a property holds
for a large class of generalized Dirac operators on an arbitrary
connected Riemannian manifold. The physical interpretation is that
single electrons cannot be localized even in curved space and the
corresponding Dirac quantum field has the Reeh-Schlieder property.
Our class of operators is large enough to allow coupling to
arbitrary smooth Hermitian potentials. Note that in this situation
only the space is curved which in a relativistic framework would
correspond to an ultrastatic space-time, \ie the space-time of the
form $\mR \times M$ with metric tensor $dt^2-g$. There already
exist results on the Reeh-Schlieder property for the free scalar
field over such space-times (\cite{verch93a}) and recently also
for quite general free quantum fields over space-times admitting a
time-like Killing vector field (\cite{strohmaier00ppa}). See also
\cite{requardt86a} for the connection of unique continuation
theorems with the Reeh-Schlieder property within non-relativistic
quantum mechanics.

The weak unique continuation property for elements of spectral
subspaces provides a particularly straightforward way to
understand the localization problem for single particles and the
Reeh-Schlieder property for the corresponding quantum fields over
ultrastatic space-times. Independently such theorems play an
important role in Riemannian geometry since splittings of the type
$\cH=\cH_+\oplus\cH_-$ also occur in an essential manner in
boundary value problems for Dirac equations
(\cite{atiyah-patodi-singer75a}). This note can be considered a
continuation of \cite{baer99ppa} since we are now able to remove
the assumption in \cite{baer99ppa} that the underlying manifold be
compact.


\section{Notation and Result}

Let $(M,g)$ be a Riemannian manifold, let $E\rightarrow M$ be a
Hermitian vector bundle over $M$. Denote the Hermitian metric on
$E$ by $\langle \cdot , \cdot \rangle$. The space of smooth
sections in $E$ will be denoted $C^{\infty}(M,E)$ and similar
notation will be used for compactly supported smooth sections and
for square-integrable sections. Let $$
D:C^{\infty}(M,E)\rightarrow C^{\infty}(M,E) $$ be a formally
self-adjoint differential operator of first order. We call $D$ of
{\it Dirac type} if its principal symbol $\sigma_D$ satisfies the
Clifford relations, \ie $$ \sigma_D
(\xi)\circ\sigma_D(\eta)+\sigma_D(\eta)\circ\sigma_D(\xi)=
-2g(\xi,\eta)\cdot\mbox{Id}_{E_p} $$ for all $\xi$, $\eta\in T^*_p
M$, $p\in M$. Then $D$ is an {elliptic} differential operator of
first order.

{\bf Example 1 } Let $M=S^1$ be the circle and let $E$ be the
trivial complex line bundle over $M$. Then $D=i\ddt$ is of Dirac
type. Fourier expansion of complex valued functions is nothing but
the eigenvector expansion for $D$.

{\bf Example 2 } More generally, any generalized Dirac operator in
the sense of Gromov and Lawson (\cite{gromov-lawson83a}) is of
Dirac type.

{\bf Example 3 } Let $M=\mR^3$ and let $D$ be the Dirac operator
in an external field, $D = H = H_0 + V$ in the terminology of
\cite[Section 4]{thaller92a}. Here the potential $V$ to be added
to  the free Dirac operator $H_0$ may be any smooth function $V$
with values in Hermitian matrices. Examples include
electromagnetic vector potentials, and electric and magnetic
anomalous moments.

{\bf Example 4 } Since the underlying manifold $M$ need not
necessarily be complete we can also deal with singular potentials.
Let us demonstrate this for the {\em Coulomb potential}. Put
$M=\mR^3-\{0\}$, $V(x):=\frac{\ga}{|x|}\Id$, and $D=H=H_0+V$ as in
the previous example. It is known (\cite[Section~4]{thaller92a})
that if the coupling constant $\ga$ is not too large, $|\ga|<
const$, then $D$ is essentially self-adjoint on
$C_c^\infty(M,\mC^4)$. But even for larger coupling constant for
which essential self-adjointness is known to break down there
still exist self-adjoint extensions and our theorem can be
applied.

\hypertarget{ex}{{\bf Example 5 }} Let $(M,g)$, $E$, and $D$ be as
above with $D$ of Dirac type. Put $X:=M\x\mR$ with Riemannian
metric $ds^2 = g + dt^2$ where $t$ is the standard coordinate on
$\mR$. Hence $X$ is the {\em cylinder} over $M$. The pull-back of
$E$ to $X$ is again denoted by $E$. Define the Hermitian vector
bundle $\cE := E\oplus E$ over $X$. Then $$ \cD := \pmatrix{D &
\delt \cr
                -\delt & -D}
$$ acting on sections in $\cE$ is of Dirac type over $X$.

Now let $D$ be of Dirac type over the \cnct \Rm \mfd $M$. Let
$\ph$ be in the kernel of $D$. By elliptic regularity theory $\ph$
is necessarily smooth. It is well-known that $D$ has the {\em
unique continuation property} meaning that if $\ph$ vanishes on a
non-empty open subset $\cO\subset M$, then $\ph$ vanishes
everywhere, see \eg \cite{booss99ppa, booss-wojciechowski93a}. Of
course this is also true for eigensections of $D$ to other
eigenvalues $\la\not= 0$ since one can simply replace $D$ by
$D-\la\Id$. This unique continuation property of Dirac type
operators can also be shown by passing to the square of $D$ and
applying the classical Aronszajn theorem (\cite{aronszajn57a}) for
Laplace type operators.

We show here that the assumption $D\ph =0$ can be relaxed
considerably. Roughly speaking, it is enough that $\ph$ lies in a
$D$-invariant subspace of the Hilbert space of square-integrable
sections $L^2(M,E)$ on which $D$  is semibounded.

For a subset $S\subset \mR$ denote the characteristic function by
$\chi_{S}$, $$ \chi_{S}(\la) = \left\{
\begin{array}{cl}
1, & \la \in S \\ 0, & \la \in \mR-S .
\end{array}
\right. $$

Our main result is

{\bf Theorem } {\em Let $M$ be a \cnct \Rm \mfd, let $E$ be a
Hermitian vector bundle over $M$, and let $D$ be of Dirac type
acting on $C_c^\infty(M,E)$. Suppose $D$ has a self-adjoint
extension in $L^2(M,E)$ which we denote by $A$. Let $\cO\subset M$
be a non-empty open subset of $M$. Let $\ph\in L^2(M,E)$ such that
there exists $\la_0\in\mR$ with $\chi_{[\la_0,\infty)}(A)\ph=0$ or
$\chi_{(-\infty,\la_0]}(A)\ph=0$.

Then the following conclusion holds: $$ \ph|_\cO = 0  \dann \ph=0
\mbox{ on all of }M. $$ }

{\bf Example 6 } If $\ph$ is a square-integrable eigensection for
$D$, then clearly, $\chi_{[\la_0,\infty)}(A)\phi=0$ for any
$\la_0$ larger than the eigenvalue and we recover the classical
unique continuation property of $D$ mentioned above.

{\bf Example 7 } In case $M=S^1$ and $D=i\ddt$ we may apply the
theorem to all $\ph\in L^2(S^1,\mC)$ whose Fourier coefficients
$c_n$ vanish for all $n$ smaller than some $n_0$, \ie all $\ph$ of
the form $$ \ph(t) = \sum_{n\ge n_0} c_n e^{int}. $$ The theorem
says that any such $\ph$ must be identically zero if it vanishes
on a non-empty open subset of $S^1$.

{\bf Example 8 } More generally, if $M$ is \cpt and \cnct, then
the spectrum of $A$ is discrete. We may split $L^2(M,E) = \cH_{\le
0} \oplus \cH_{>0}$ where $\cH_{\le 0}$ denotes the sum of all
eigenspaces for the non-positive eigenvalues and similarly
$\cH_{>0}$ corresponds to the positive eigenvalues. This splitting
is important to define the Atiyah-Patodi-Singer boundary
conditions for boundary value problems for Dirac operators
(\cite{atiyah-patodi-singer75a}). By the theorem no non-trivial
$\ph\in \cH_{\le 0}$ or $\ph\in \cH_{> 0}$ can vanish on a
non-empty open subset of $M$.

{\bf Remark } Note that in the theorem no assumption of essential
self-adjointness of $D$ on $C^\infty_c(M,E)$ is made. Dirac type
operators on a {\em complete} manifold are always essentially
self-adjoint on smooth compactly supported sections. The
well-known proofs of this fact for  generalized Dirac operators in
sense of Gromov and Lawson (\cite{gromov-lawson83a}) easily carry
over to our more general class of operators, see also
\cite{chernoff73a} for a different approach. However, we only need
existence of self-adjoint extensions not their uniqueness. Hence
in our case the manifold $M$ need not necessarily be complete,
compare Example 4. Also see \cite{lesch93a} for examples with cone
singularities to which our theorem can be applied.

It should also be emphasized that $\ph$ need not lie in the domain
of definition of $D$ or of $A$. Only square-integrability and
$\chi_{[\la_0,\infty)}(A)\ph=0$ or
$\chi_{(-\infty,\la_0]}(A)\ph=0$ are assumed.

Also note that our theorem shows that the spectrum of any Dirac
type operator is indeed unbounded from above and from below.
Otherwise, every section in $L^2(M,E)$ would have the unique
continuation property which is absurd.


\section{The Proof}

In this section we give the proof of the theorem. We keep the
notation of the theorem and we assume $\ph|_\cO = 0$. We wish to
show $\ph=0$.

Without loss of generality we may assume that $\la_0=1$ and that
$\chi_{(-\infty,1]}(A)\ph=0$. In other words, $\ph$ lies in the
image of the projection $\chi_{(1,\infty)}(A)$. Denote the upper
half plane by $$ \goth{H} := \{ \ze\in\mC\ |\ \Im(\ze) >0 \}. $$

{\em Step 1: } The family of functions $f_z$, $$ f_z(\la) =
\left\{
\begin{array}{cl}
\la\inv e^{iz\la}, & \la\ge 1\\ 0,         & \la < 1
\end{array}
\right. , $$ is uniformly bounded by 1 for all
$z\in\bar{\goth{H}}$ and for each fixed $\la$ it is continuous in
$z$. Therefore the family of bounded operators $f_z(A)$ is
continuous in $z\in\bar{\goth{H}}$ in the strong operator
topology. Hence $z\mapsto f_z(A)\ph$ is a continuous
$L^2(M,E)$-valued function on $\bar{\goth{H}}$.

Now fix $z_0\in\goth{H}$. The family of functions $g_z$ defined by
$$ g_z(\la) := \left\{
\begin{array}{cl}
\la^{-1}\frac{e^{iz\la}-e^{iz_0\la}}{z-z_0}, & z\not= z_0\\
ie^{iz_0\la}, & z=z_0
\end{array}
\right. $$ for $\la\ge 1$ and $g_z(\la)=0$ for $\la < 1$ is also
uniformly bounded and continuous in $z$. This shows $$ \lim_{z\to
z_0}\frac{f_z(A)\ph - f_{z_0}(A)\ph}{z-z_0} $$ exists and hence
$z\mapsto f_z(A)\ph$ is holomorphic on $\goth{H}$.

{\em Step 2: } Fix a non-empty relatively compact open subset
$\tilde{\cO} \subset\subset \cO$. Let $u\in C^\infty_c(M,E)$ with
support contained in $\tilde{\cO}$. By finiteness of propagation
speed, see \hyperlink{speedy}{Section} \ref{speed}, there exists
$\ep >0$ such that the support of $e^{itA}u$ is contained in $\cO$
for all $t\in [-\ep,\ep]$. Hence for all $t$ with $|t|\le\ep$ we
have
\begin{eqnarray*}
0 &=& (\ph,e^{-itA}u)\\ &=& (e^{itA}\ph,u) \\ &=&
(Af_{t}(A)\ph,u)\\ &=& (f_{t}(A)\ph,Au) .
\end{eqnarray*}
Here we used that by assumption $\ph$ lies in the image of the
projection $\chi_{(1,\infty)}(A)$ and hence
$e^{izA}\ph=Af_{z}(A)\ph$. Since $z\mapsto (f_{z}(A)\ph,Au)$ is
continuous on $\bar{\goth{H}}$ and holomorphic on $\goth{H}$ and
vanishes on $[-\ep,\ep]$ an application of the Schwarz reflection
principle (\cite[Lemma~2]{baer99ppa}) shows $$ (f_{z}(A)\ph,Au) =
0 $$ for all $z\in\bar{\goth{H}}$. In particular for $z=it$ with
$t>0$ we obtain $$ (e^{-tA}\ph,u) = (f_{it}(A)\ph,Au) = 0 $$ for
all $u\in C^\infty_c(M,E)$ with support contained in
$\tilde{\cO}$. Hence $e^{-tA}\ph$ vanishes on $\tilde{\cO}$ for
all $t>0$.

{\em Step 3: } Now look at the half cylinder $X=M\x (0,\infty)$
with product metric $g+dt^2$, Hermitian vector bundle $\cE :=
E\oplus E$ over $X$, and Dirac type operator $$ \cD := \pmatrix{D
& \delt \cr
                -\delt & -D}
$$ as in \hyperlink{ex}{Example 5}. We define a distributional
section $\Phi_1$ in $E$ over $X$ by $$ \Phi_1(v) := \int_0^\infty
(e^{-tA}\ph, v(\cdot,t))_{L^2(M,E)}dt $$ for all $v\in
C_c^\infty(X,E)$. The differential equation $$ \ddt e^{-tA}\ph = -
Ae^{-tA}\ph $$ shows that the distributional section $$ \Phi :=
\pmatrix{\Phi_1 \cr \Phi_1} $$ in $\cE$ satisfies $$ \cD\Phi =0 $$
in the distributional sense. In particular, $\Phi$ is smooth by
elliptic regularity theory.

Since $\Phi$ vanishes on the open subset $\tilde{\cO}\x
(0,\infty)$ of $X$ the standard unique continuation property of
$\cD$ implies $\Phi=0$. Hence $\Phi_1=0$, \ie $e^{-tA}\ph=0$ for
all $t>0$, and the limit $t\searrow 0$ yields $\ph=0$. \qed


\section{Finite Propagation Speed}
\label{speed}

\hypertarget{speedy}{In }the second step of the proof we needed
what is known as ``finite propagation speed'' for possibly
incomplete manifolds. For the sake of completeness we give a
justification for this by reducing it to the compact case. A
direct approach can be found in \cite{chernoff73a}.

For any closed subset $A$ of $M$ let $U_r(A)$ be the closed
$r$-neighborhood of $A$, \ie $$ U_r(A) = \{ x\in M\ |\ \dist(x,A)
\le r \}. $$

{\bf Proposition } {\em Let $M$ be a Riemannian manifold, let $E$
be a Hermitian vector bundle over $M$, and let $D$ be of Dirac
type acting on $C_c^\infty(M,E)$. Suppose $D$ has a self-adjoint
extension in $L^2(M,E)$ which we denote $A$. Let $\ph\in
C^\infty_c(M,E)$.

Then there exists $\ep>0$ such that for all $t\in [0,\ep)$ $$
\ph_t := e^{itA}\ph $$ is smooth in all variables and $$
\supp(\ph_t) \subset U_t(\supp(\ph)). $$ If $M$ is complete, then
the statement is true for $\ep=\infty$. }

\pf The proposition is well-known if the underlying manifold is
compact, see \eg \cite[Prop.~5.5]{roe88a} for a proof. By
assumption $\supp(\ph)$ is compact. Hence there exists $\ep>0$
such that $U_\ep(\supp(\ph))$ is still compact. We pick a compact
manifold $Y$ containing an isometric image of $U_\ep(\supp(\ph))$
together with a Hermitian bundle and Dirac type operator extending
$E$ and $D$ over $U_\ep(\supp(\ph))$. (This can \eg be obtained by
choosing an open subset $X\subset M$ containing
$U_\ep(\supp(\ph))$ with compact closure $\bar{X}$ and smooth
boundary $\del X$. Then let $Y$ be the double of $X$ and smooth
out all data along $\del X$.) Since $\ph$ has support contained in
$U_\ep(\supp(\ph))$ we can consider it as a section over $Y$ by
extending it by $0$. Let $\ph_t$ denote the solution of $$ \delt
\ph_t = iD\ph_t $$ over $Y$ with $\ph_0=\ph$. Then $\ph_t$ is
smooth in all variables and has support contained in
$U_\ep(\supp(\ph))$ for $t\in [0,\ep)$. By this last property we
can consider $\ph_t$ as a smooth section over $M$, again extending
by zero. Considered as elements of $L^2(M,E)$ we have $$ \ddt
\ph_t = iD\ph_t = iA\ph_t $$ and hence $\ph_t=e^{itA}\ph$.

In case $M$ is complete $\ep$ can be chosen arbitrarily large.
\qed


\section{Concluding Remarks}

Our main argument makes use of the unique continuation theorem for
solutions to Dirac type equations on Riemannian manifolds (on the
half cylinder $X$). In order to use this theorem we had to
analytically continue the functions $t\mapsto (f_{t}(A)\ph,Au)$ (see Step
2 in the proof) to ``imaginary time''. This was possible due to
semi-boundedness of the spectrum of the Dirac operator on the
spectral subspaces under consideration. In quantum field theory
this method is commonly referred to as ``Wick rotation''.

It should be noted that our theorem can be extended to even more
general subspaces. Clearly analyticity of the function
$(f_{z}(A)\ph,Au)$ in the whole upper half plane is not necessary,
but it suffices to have boundedness and analyticity in a strip of
the form $\{ \ze\in\mC\ |\ 0 < \Im(\ze) < \beta \}$ for some
$\beta >0$. Such subspaces arise naturally in thermal quantum
field theory.


\providecommand{\bysame}{\leavevmode\hbox to3em{\hrulefill}\thinspace}

\vspace{0.5cm}

\parskip0ex

Fachbereich Mathematik

Universit\"at Hamburg

Bundesstr.~55

20146 Hamburg

Germany

\vspace{0.4cm}

E-Mail: {\tt baer@math.uni-hamburg.de}

WWW: {\tt http://www.math.uni-hamburg.de/home/baer/}

\vspace{0.4cm}

and

\vspace{0.4cm}

Institut f\"ur theoretische Physik

Universit\"at Leipzig

Augustusplatz 10/11

04109 Leipzig

Germany

\vspace{0.4cm}

E-Mail: {\tt alexander.strohmaier@itp.uni-leipzig.de}

WWW: {\tt http://www.physik.uni-leipzig.de/\~{}strohmai/}
\end{document}